\documentclass[prl,nofootinbib,preprintnumbers,tightenlines,twocolumn,superscriptaddress]{revtex4-2}
\usepackage{braket}
\usepackage{array}
\usepackage{dsfont}
\usepackage{amsmath}
\usepackage{pifont}
\usepackage[normalem]{ulem}
\usepackage{amssymb}
\usepackage{xcolor}
\usepackage{amsfonts}
\usepackage{mathtools}
\usepackage{dcolumn}
\usepackage{bm}
\usepackage{multirow}
\usepackage{hhline}
\usepackage[pdftex]{graphicx,hyperref}
\usepackage{leftidx}
\usepackage{color}
\usepackage{mathtools}
\usepackage{MnSymbol}
\usepackage[mathscr]{eucal}
\usepackage[german,english]{babel}
\usepackage[capitalize]{cleveref}

\usepackage[utf8]{inputenc}
\usepackage{natbib}
\usepackage{comment}

\begin{document}

\title{Theory of broken symmetry quantum Hall states in the $N=1$ Landau level of Graphene}
\author{Nikolaos Stefanidis}
\email{nikos@pks.mpg.de}
\affiliation{Max-Planck-Institut für Physik komplexer Systeme, Dresden 01187, Germany}
\author{Inti Sodemann Villadiego}
\email{sodemann@itp.uni-leipzig.de} 
\affiliation{Institut für Theoretische Physik, Universität Leipzig, D-04103, Leipzig, Germany}
\affiliation{Max-Planck-Institut für Physik komplexer Systeme, Dresden 01187, Germany}

\begin{abstract}
We study many-body ground states for the partial integer fillings of the $N=1$ Landau level in graphene, by constructing a model that accounts for the lattice scale corrections to the Coulomb interactions. Interestingly, in contrast to the $N=0$ Landau level, this model contains not only pure delta function interactions but also some of its derivatives. Due to this we find several important differences with respect to the $N=0$ Landau level. For example at quarter filling when only a single component is filled, there is a degeneracy lifting of the quantum hall ferromagnets and ground states with entangled spin and valley degrees of freedom can become favourable. Moreover at half-filling of the $N=1$ Landau level, we have found a new phase that is absent in the $N=0$ Landau level,  that combines  characteristics of the Kekul\'{e} state and an antiferromagnet. We also find that according to the parameters extracted in a recent experiment, at half-filling of the $N=1$ Landau level graphene is expected to be in a delicate competition between an AF and a CDW state, but we also discuss why the models for these recent experiments might be missing some important terms.
\end{abstract}

\maketitle

\noindent

{\it \textcolor{blue}{Introduction.}} The quantum Hall regime in graphene realizes a rich landscape of broken symmetry and topological states, stemming in part from the near four-fold degeneracy of its Landau levels (LLs) associated with its valley and spin degrees of freedom~\cite{goerbig2011electron}. Most studies to date have focused on the states in the $N=0$ LL, with transport and magnon transmission experiments favoring an anti-ferromagnetic (AF) state at neutrality~\cite{young2014tunable,wei2018electrical,stepanov2018long,zhou2022strong,paul2022electrically}, while STM experiments reporting evidence for Kekul\'{e}-type valence-bond-solids and charge density wave states (CDW)~\cite{li2019scanning,liu2022visualizing,coissard2022imaging}.
\par While the projected Coulomb interaction is typically the dominant term in the Hamiltonian, it possesses a large symmetry that leaves the quantum Hall ground states undetermined. Therefore, it is crucial to account for the corrections that reflect the lower symmetry of the underlying graphene lattice to select the ground states~\cite{goerbig2011electron,nomura2006quantum,goerbig2006electron,alicea2006graphene,herbut2007theory,jung2009theory,kharitonov2012phase}. A convenient model to capture these symmetry breaking interactions in the $N=0$ LL was introduced by Kharitonov in Ref.~\cite{kharitonov2012phase}. This model can be viewed as a projection into the $N=0$ LL of a more general model introduced by Aleiner, Kharzeev and Tsvelik~\cite{aleiner2007spontaneous, kharitonov2012phase} that includes all possible delta-function interactions allowed by symmetries. There is no study to this date that has constructed an analogous model in the $N=1$ LL that includes all possible short-distance interactions allowed by symmetry, although a related model containing some of these terms was introduced in Ref.~\cite{yang2021experimental}.
\par The purpose of our study is therefore to construct this model of symmetry breaking interactions in the $N=1$ LL and to determine its ground states at partial integer fillings. Intrestingly, we will see that in contrast to the $N=0$ LL~\cite{kharitonov2012phase}, the $N=1$ LL model contains interactions that are not pure delta functions~\cite{yang2021experimental}. Therefore, in contrast to the $N=0$ LL, a unique ground state is selected even when a single component is filled (to be denoted by $\tilde{\nu}=1$), and some of the possible ground states are spin-valley entangled, in the sense discussed in Ref.~\cite{atteia20214}. Moreover, when two components are filled (to be denoted by $\tilde{\nu}=2$), we find a new type of  Kekul\'{e}-Antiferromagnetic state in addition to those found in the $N=0$ LL. Based on the parameters estimated in Ref.~\cite{yang2021experimental}, graphene is expected to be in a delicate competition between an AF and a CDW state. However, as we will discuss, these parameters are possibly missing some important terms.

\noindent

{\it \textcolor{blue}{Model and Symmetries.}} 
We begin by reviewing the continuum model of short-range symmetry breaking interactions of Aleiner, Kharzeev and Tsvelik~\cite{aleiner2007spontaneous} in the absence of a magnetic field. This is described by the following Hamiltonian:
\begin{equation} \label{general}
\mathcal{H}= \mathcal{H}_D+\mathcal{H}_C+\mathcal{H}_A ,
\end{equation}
with: 
\begin{equation} \label{Dirac}
\mathcal{H}_D=v_F \sum_{i} (\tau^{i}_z p^{i}_x \sigma^{i}_x+p^{i}_y \sigma^{i}_y),
\end{equation}
being the linearized single particle hamiltonian around the Dirac points, 
$$
\mathcal{H}_C=\sum_{i<j}\frac{e^2}{\epsilon |\mathbf{r}_i-\mathbf{r}_j|},
$$
the Coulomb interaction, and 
\begin{equation} \label{Asymmetry}
\mathcal{H}_A= \sum_{i<j}\{ \sum_{\alpha,\beta}V_{\alpha \beta} T^{i}_{\alpha \beta} T^{j}_{\alpha \beta}\} \delta(\mathbf{r}_i-\mathbf{r}_j) ,
\end{equation}
the sublattice-valley dependent interactions. We have defined $T^{i(j)}_{\alpha \beta}=\tau_{\alpha}^{i(j)} \otimes \sigma_{\beta}^{i(j)}\otimes s_{0}^{i(j)}$, and $\tau_{\alpha}^{i(j)}, \ \sigma_{\beta}^{i(j)}, \ s_{0}^{i(j)}$ $\alpha,\beta=0,x,y,z$ to be the Pauli matrices acting on valley, sublattice and spin respectively.
\par
\begin{figure}[t!]
\centering
\includegraphics[width=0.48\textwidth]{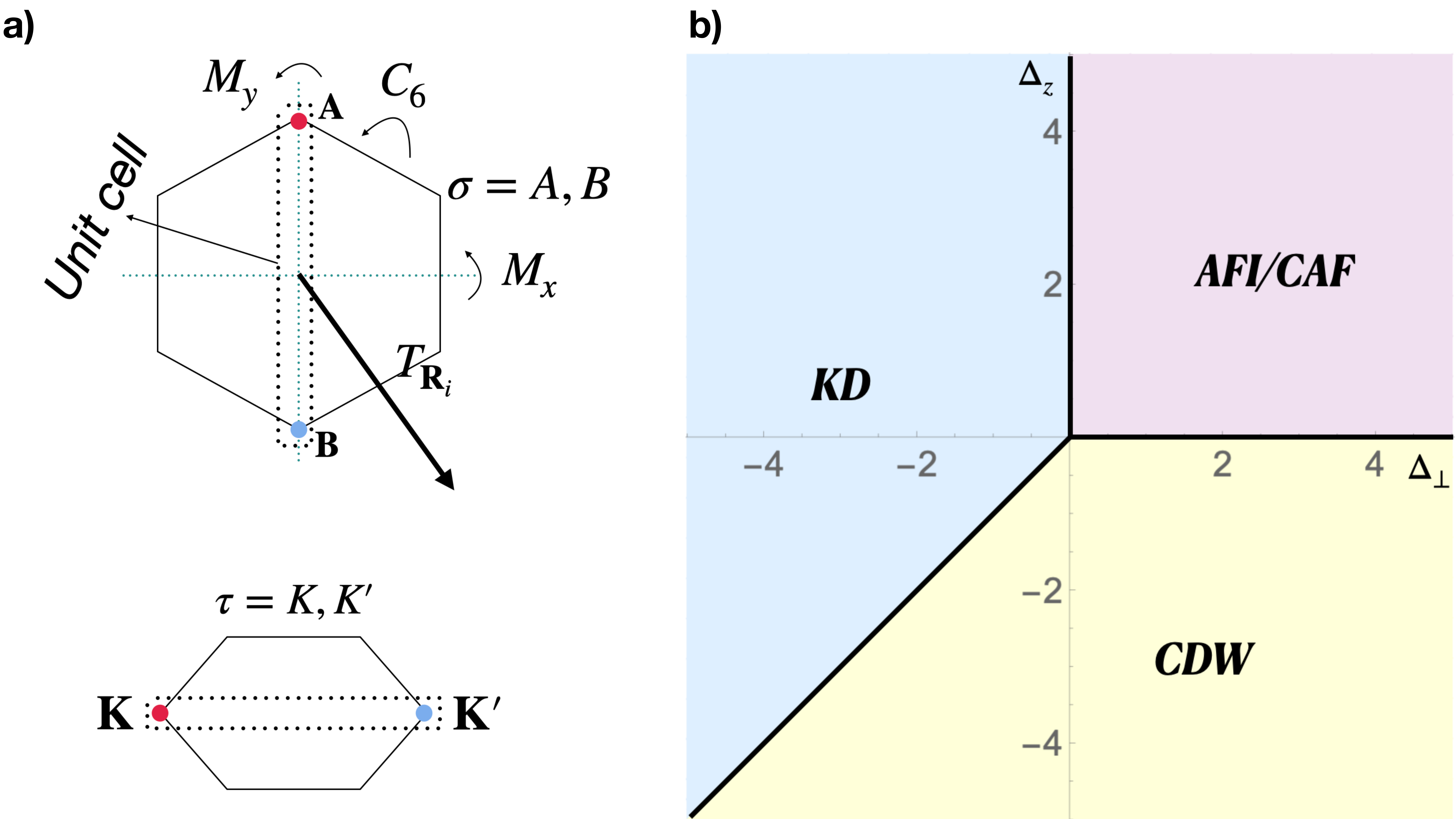}   
\caption{a) Graphene unit cell and its lattice symmetries (top) and its reciprocal unit cell (bottom). b) Phase diagram at $\tilde{\nu}=1$. It contains four phases: charge density wave (CDW), Kekul\'{e} distortion (KD) and the two entangled phases, the antiferrimagnetic phase (AFI) and the canted antiferromagnet (CAF).}
\label{Goerbigs}  
\end{figure}

 By denoting the valley (sublattice) states as $\ket{\tau}(\ket{\sigma})$, with $\tau(\sigma) = \pm 1$ corresponding to the $K,K' \ (A,B)$ valleys (sublattices), then the action of lattice symmetry on these states is given by:
\begin{equation} \label{symm ops}
\begin{split}
&C_{6}\ket{\tau,\sigma}=Z^{-\tau \sigma}\ket{-\tau,-\sigma}, \\  
&M_{x}\ket{\tau,\sigma}=\ket{\tau,-\sigma}, \\
&M_{y}\ket{\tau,\sigma}=\ket{-\tau,\sigma},\\
&T_{R_{1,2}} \ket{\tau,\sigma}=Z^{\pm \tau } \ket{\tau, \sigma},\\
\end{split}
\end{equation}
with $Z=e^{i \frac{2 \pi}{3}}$. $C_{6}$ is the rotation by $\pi/3$, $M_{x}, M_{y}$ the two mirrors and $T_{R_i}$ the translations by the two basis vectors of graphene (see Fig.~\ref{Goerbigs}(a) for the illustration of these symmetries). These symmetries reduce the couplings of Eq.~\eqref{Asymmetry} to nine independent couplings satisfying the following relations~\cite{aleiner2007spontaneous}:
\begin{equation} \label{symmetries}
\begin{split}
&F_{\perp z}  \equiv V_{xx}=V_{yx}, \\
&F_{z \perp}\equiv V_{0x}=V_{zy}, \\
&F_{\perp \perp}\equiv V_{xz}=V_{y0}=V_{yz}=V_{x0}, \\
&F_{0 \perp}\equiv V_{zx}=V_{0y}, \\
&F_{\perp 0}\equiv V_{yy}=V_{xy}, \\
&F_{zz}\equiv V_{0z}, \\
&F_{z0}\equiv V_{z0}, \\
&F_{0z}\equiv V_{zz} . \\
\end{split}
\end{equation}
\noindent
{\it \textcolor{blue}{Projected Model in the $N=1$ Landau level.}}
 By projecting $\mathcal{H}_A$ from Eq.~\eqref{general}, with the constraints in Eq.~\eqref{symmetries}, one obtains the following Hamiltonian of symmetry breaking interactions in the $N$th LL:
\begin{equation} \label{proj model}
\mathcal{H}^{N}_{A}=\sum_{i<j} \{ V^{N}_{z}(r_{ij}) \tau^{i}_z\tau^{j}_z + V^{N} _{\perp}(r_{ij})\tau^{i}_{\perp}\tau^{j}_{\perp}\},
\end{equation}
with $\tau^{i}_{\perp}\tau^{j}_{\perp}=\tau^{i}_{x}\tau^{j}_{x}+\tau^{i}_{y}\tau^{j}_{y}$ (see S-II for further details). As we see there is an effective $U(1)$ valley conservation arising from the underlying lattice symmetries. Specifically for the $N=1$ LL we have: 
\begin{equation} \label{strengths N=1}
        V_{z,\perp}(r_{ij}) = \sum_{n=0}^{2}g^{z,\perp}_n \nabla^{2n} \delta(r_{ij}).\\
\end{equation}
Here $g^{z,\perp}_n$ are independent constants that parametrize the projected interactions that are linear combinations of those in Eq.~\eqref{symmetries} (see Eq.(S-28) for their explicit relations). Therefore we have a model with $6$ independent parameters characterizing the interactions in the $N=1$ LL, in contrast to the more restricted model of Ref.~\cite{yang2021experimental} with only 2 parameters. The model of Ref.~\cite{yang2021experimental} is a special case of our Eq.~\eqref{strengths N=1}, in which $g^{z}_{0,1}=g^{\perp}_{0,2}=0$. Notice, in particular, that in our model the $n=0$ terms in Eq.~\eqref{strengths N=1} are pure delta function interactions, which are absent in Ref.~\cite{yang2021experimental} (see S-IV for further details).
\par On the other hand, if we project $\mathcal{H}_A$ onto the $N=0$ LL we obtain the model from Ref.~\cite{kharitonov2012phase} for which the interactions would include only pure delta functions (see Eq.(S-29) for the definition of $g_{z, \perp}$):
\begin{equation} \label{strengths N=0}
       V_{z, \perp}(r_{ij})= g_{z, \perp} \delta(r_{ij}). \\ 
\end{equation}
Therefore, the main difference between the model of Eq.~\eqref{strengths N=1} for the $N=1$ LL and the model of Ref.~\cite{kharitonov2012phase} for the $N=0$ LL is the existence of interactions which are not pure delta functions. As we will show, this leads to several important differences in the physics of these two Landau levels.

{\it \textcolor{blue}{Mean-field ground states.}} We will now derive the Hartree-Fock (HF) functional for the Hamiltonian of Eqs.~\eqref{proj model},~\eqref{strengths N=1} and obtain the phase diagram in the integer fillings of the $N=1$ LL, $\tilde{\nu}=1$ ($\tilde{\nu}=2$) when one (two) out of the four valley-spin degenerate LL are filled \footnote{The partial filling of $\tilde{\nu}=3 (\tilde{\nu}=2)$ is equivalent to $\tilde{\nu}=1(\tilde{\nu}=4)$ by a particle-hole conjugation.}. We consider the competition of translational invariant integer quantum Hall ferromagnets that can be described by a particle-hole condensate order parameter of the form, $\braket{c^{\dagger}_{X_1 \tau_1 s_1} c_{X_2 \tau_2 s_2}}=P^{s_1 s_2}_{\tau_1 \tau_2} \delta_{X_1,X_2}$, with $X_i$ labeling intra-LL guiding center coordinates. Here $c^{\dagger}_{X\tau s}$ denotes the electron creation operator with valley $\tau$ and spin $s$, and $P$ is the projector in spin-valley space into either a one-dimensional subspace (for $\tilde{\nu}=1$) or a two-dimensional subspace (for $\tilde{\nu}=2$). The general form of the Hartree-Fock functional is then ($\mathcal{E}_{HF}[P]  \equiv \frac{2 A}{N^2_{\phi}}E_{HF}[P]$): 
\begin{equation} \label{HF energy}
\mathcal{E}_{HF}[P] = \sum_{i=x,y,z}\bigg( u^{H}_{i} (Tr \{ T_{i} P\})^2-u^{X}_{i}Tr \{ (T_{i} P)^2\} \bigg) ,
\end{equation}
with $u^{H,X}_{\perp}=u^{H,X}_{x}=u^{H,X}_{y}$.
Therefore the possible ground states depend only on $4$ effective Hartree and exchange constants, $u^{H}_{z},u^{X}_{z},u^{H}_{\perp},u^{X}_{\perp}$, which are linear combinations of the constants $g_n^{z,\perp}$ that appear in Eq.~\eqref{strengths N=1} (see Eq.(S-35) for explicit relations). Moreover, while in the $N=0$ LL (see Eq.(S-33)) the Hartree and the exchange constants are forced to be equal, $u^{H}_{z,\perp}=u^{X}_{z,\perp}$ ~\cite{kharitonov2012phase}, in the $N=1$ LL they are independent due to the appearance of non-delta interactions (see S-III for further details). Similar functionals have been proposed, however phenomenologically,  for the $N=0$ LL to capture the physics beyond the delta-functions in Refs.~\cite{atteia20214,das2022coexistence}. 
\par We will consider general spin-valley entangled ~\cite{atteia20214,das2022coexistence} variational states. The following two orthonormal spinors can be used to uniquely parametrize the state characterized by $P$ in Eq.~\eqref{HF energy},
\begin{equation} \label{entangled states}
\begin{split}
\ket{F}_1&=\cos\frac{a_1}{2} \ket{\boldsymbol{\eta}}\ket{\mathbf{s}}+e^{i \beta_1} \sin \frac{a_1}{2} \ket{-\boldsymbol{\eta}} \ket{-\mathbf{s}}, \\
\ket{F}_2&=\cos\frac{a_2}{2} \ket{\boldsymbol{\eta}}\ket{\mathbf{-s}}+e^{i \beta_2} \sin \frac{a_2}{2} \ket{-\boldsymbol{\eta}} \ket{\mathbf{s}}.\\
\end{split}
\end{equation} 
Here  $\ket{\boldsymbol{\eta}}$ and $\ket{\mathbf{s}}$ are states parametrized by unit vectors $\boldsymbol{\eta}$ and $\mathbf{s}$ in the spin and valley Bloch spheres respectively and $a_{1,2}$ and $\beta_{1,2}$ are real constants. Notice that in general these states might not be separable into a tensor product of spin and valley components and therefore can account for spin-valley entanglement~\cite{atteia20214}. For $\tilde{\nu}=1$, we take $P=| F_1><F_1 |$, and for $\tilde\nu=2$ $P=| F_1><F_1 |+| F_2><F_2 |$.

{\it \textcolor{blue}{Ground states for $\tilde{\nu}=1$.}}
\noindent
As discussed in Ref.~\cite{atteia20214}, the energy functional in this case reduces to:
\begin{equation} \label{HF energy-first}
\mathcal{E}_{HF}^{\tilde{\nu}=1}=\cos^2 a_1 \{ \Delta_z \eta_z^2+\Delta_{\perp} \eta_{\perp}^2\},
\end{equation}
with $\Delta_z=u_{z}^{H}-u_{z}^{X}$, $\Delta_{\perp}=u_{\perp}^{H}-u_{\perp}^{X}$ and $\eta_{\perp}^2=\eta_{x}^2+\eta_{y}^2$ (see S-V-A) for further details). The resulting phase diagram is shown in Fig.\ref{Goerbigs}(b) and contains four phases. These are a charge density wave (CDW) with $\boldsymbol{\eta}=\hat{z}, \ \boldsymbol{s}=\hat{z}$ and $a_1=0$, and a Kekul\'{e} distortion (KD) state with $\boldsymbol{\eta}=\boldsymbol{\eta}_{\perp}, \ \boldsymbol{s}=\hat{z}$ and $a_1=0$. Interestingly, we see that also spin-valley entangled phases with $a_1=\pi/2$, appear when $\Delta_z>0, \  \Delta_{\perp}>0$. These entangled phases are degenerate in the absence of Zeeman fields, but in their presence they split antiferrimagnetic phase (AFI) with $\boldsymbol{\eta}=\hat{z}, \ \boldsymbol{s}=\hat{z}$ and $a_1=\frac{\pi}{2}$ and the canted antiferromagnet (CAF) with $\boldsymbol{\eta}=\boldsymbol{\eta}_{\perp}, \ \boldsymbol{s}=\hat{z}$ and $a_1=\frac{\pi}{2}$, as discussed in Ref.~\cite{atteia20214}.

\par Notice that in the $N=0$ LL, $\Delta_z=\Delta_{\perp}=0$, and therefore all of the above states would be degenerate and with a vanishing HF energy.

{\it \textcolor{blue}{Ground states for $\tilde{\nu}=2$.}} The HF functional for $\nu=2$ is more difficult to minimize analytically. To make progress, we first consider the subset of states from Eq.~\eqref{entangled states} without spin-valley entanglement. These can be classified into the valley active states~\cite{hegde2022theory} : 
\begin{equation} \label{valley active states}
\ket{F}_1= \ket{\boldsymbol{\eta}_1}\ket{\mathbf{s}}, \ 
\ket{F}_2= \ket{\boldsymbol{\eta_2}}\ket{-\mathbf{s}} ,
\end{equation}
in which the valley degree of freedom varies, and the spin active states :
\begin{equation} \label{spin active states}
\ket{F}_1= \ket{\boldsymbol{\eta}}\ket{\mathbf{s}_1}, \ 
\ket{F}_2= \ket{\boldsymbol{-\eta}}\ket{\mathbf{s}_2} ,
\end{equation}
in which the spin degree of freedom varies.
We first minimize the energy functional within this subspace and then perform a quadratic expansion of all possible deviations of parameters that account for spin-valley entangled states (see S-V-B), VI, VII for further details). For simplicity we will also neglect the Zeeman term that is typically weak compared to the interaction terms~\cite{young2014tunable,abanin2013fractional,sodemann2014broken}. In contrast to $\tilde{\nu}=1$, for $\tilde{\nu}=2$ we find that whenever a spin-valley disentangled state is energetically favorable it is also an exact local minima of the energy with respect to all possible quadratic deviations that include spin-valley entanglement. This indicates that these spin-valley disentangled states are also possibly exact global minima of the energy. 

\begin{figure}[t!] 
     \centering  
     \includegraphics[width=0.48\textwidth]{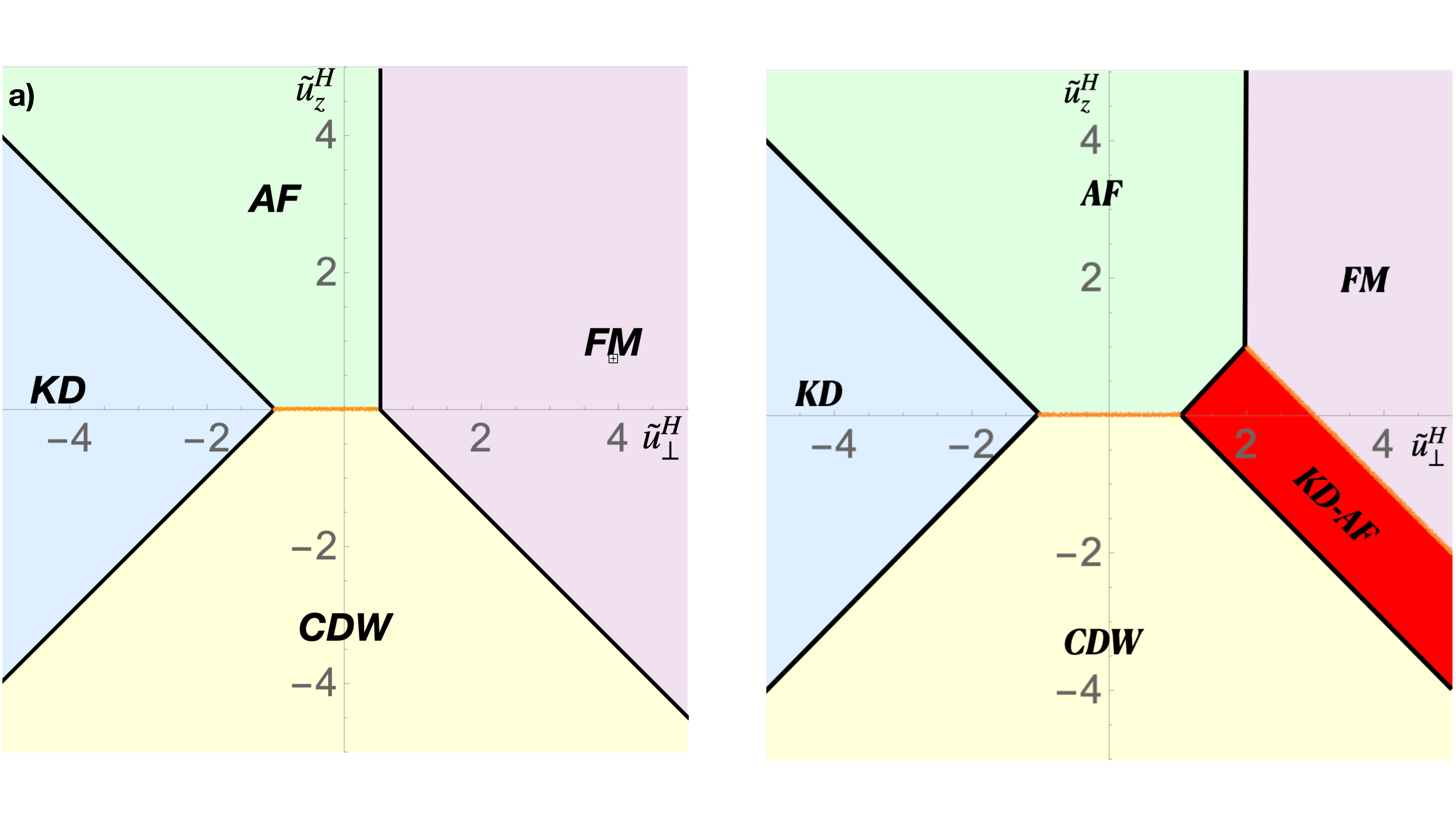}
     \caption{a) Phase diagram at $\tilde{\nu}=2$ when $\Delta_z <0 , ~ \Delta_{\perp}>0, ~ \Delta_z< \Delta_{\perp}$ for $\Delta_z/\Delta_{\perp}=-1$. According to the estimates of ~\cite{yang2021experimental} (see S-II for further details), graphene is located at the dot at the origin and therefore at the boundary between the CDW and AF phases. b) Phase diagram at $\tilde{\nu}=2$ when $\Delta_z, ~ \Delta_{\perp}>0, ~ \Delta_z< \Delta_{\perp}$ for $\Delta_z/\Delta_{\perp}=1/2$. This contains a new phase, the Kekul\'{e} distortion antiferromagnet (KD-AF), which does not appear in the $N=0$ LL. The thick black boundaries represent special first order transitions (phases become unstable coincidentally with their energy crossing) while the orange  ones indicate ordinary first order transitions (energies cross but phases remain metastable).}
     \label{Phasediagrams}
\end{figure}

Following this procedure, we find a total of five possible ground states for $\tilde{\nu}=2$ that are realized as a function of the four Hartree and exchange parameters $u^{H}_{z},u^{X}_{z},u^{H}_{\perp},u^{X}_{\perp}$. These possible five states are listed in Table~\ref{Table1} (see S-V-B) for more details on these states). To visualize the energetic competition among these five phases, we have chosen to draw two-dimensional phase diagrams as functions of the two Hartree parameters $\tilde{u}_{z,\perp}^H=u_{z,\perp}^H/|\Delta_z-\Delta_{\perp}|$ for fixed values of  $\Delta_z=u_{z}^{H}-u_{z}^{X}$, $\Delta_{\perp}=u_{\perp}^{H}-u_{\perp}^{X}$. We find that there are a total of six different kinds of phase diagrams depending on the values and signs of $\Delta_{z,\perp}$.
Two of these representative phase diagrams are depicted in Fig.~\ref{Phasediagrams}, and the remainder are presented in S-V-B).
\begin{table}[t]
\begin{tabular}{ |p{4cm}||p{4cm}|} 
 \hline
 \multicolumn{2}{|c|}{States appearing at $\tilde{\nu}=2$} \\
 \hline
States & Wavefunctions $\{ \ket{F}_1, \ket{F}_2$ \} \\
 \hline
 CDW (Charge density wave)  & $ \{\ket{\hat{z}} \ \ket{\mathbf{s}}, \ \ket{\hat{z}} \ \ket{-\mathbf{s}} \} $ \\
 KD (Kekul\'{e} distortion)& $\{ \ket{\boldsymbol{\eta}_{\perp}} \  \ket{\mathbf{s}}, \ \ket{\boldsymbol{\eta}_{\perp}} \ \ket{-\mathbf{s}} \} $  \\
 FM (Ferromagnet)& $\{\ket{\hat{z}} \ \ket{\mathbf{s}}, \ \ket{-\hat{z}} \ \ket{\mathbf{s}}\}$  \\
 AF  (Antiferromagnet)  & $\{ \ket{\hat{z}} \ket{\mathbf{s}}, \ \ket{-\hat{z}} \ket{-\mathbf{s}}$ \\
 KD-AF (Kekul\'{e} antiferromagnet)& $ \{ \ket{\boldsymbol{\eta}_{\perp}} \ket{\mathbf{s}}, \ \ket{-\boldsymbol{\eta}_{\perp}} \ket{-\mathbf{s}} \}$ \\
 \hline
\end{tabular}
\caption{Competing states at $\tilde{\nu}=2$ and their wavefunctions.}
\label{Table1}
\end{table}

Interestingly, according to the model and the estimates of Ref.~\cite{yang2021experimental}, $u^{H}_{z}=u^{H}_{\perp}=0$ and $u^{X}_{z}>0, \ u^{X}_{\perp}<0$ (see S-IV for further details). This means that graphene in the $N=1$ LL would have a phase diagram like the one in Fig.~\ref{Phasediagrams}(a), and it would be located exactly at the origin of this phase diagram, which we indicate by a black dot in Fig.~\ref{Phasediagrams}(a). Therefore, we see that the model and the parameter estimates of Ref.~\cite{yang2021experimental} place graphene right at the boundary between the CDW and the AF states. We note that even at this boundary, these phases remain stable against spin-valley entangled rotations (see S-VII for further details). 
     
\par One of the interesting qualitative differences that we have found in the $N=1$ LL is the existence of a new phase that features a combination of Kekul\'{e} state and antiferromagnet, that we term the Kekul\'{e}- antiferromagnet (KD-AF). In this phase one set of electrons has an XY vector in the valley sphere with spin up while the others occupy the opposite valley vector with spin down, as described in Table~\ref{Table1}. This phase occupies the red region in Fig.~\ref{Phasediagrams}(b).

\par In Figs.~\ref{Phasediagrams}(a)-(b) the phase transitions represented by black thick lines are a special type of first order transitions, in the sense that at these lines the energy of two states is the same, and also the quadratic expansion around them indicates an instability (or, in other words, the states do not remain metastable upon crossing this line). This makes these boundaries interesting as they are expected to be highly sensitive to perturbations which could lead to new phases or phase coexistence, as discussed in Ref.~\cite{das2022coexistence}, and also they could harbor larger symmetries, as in the $SO(5)$ symmetry in AF-Kekul\'{e} transition found in the $N=0$ LL~\cite{wu2014so}. The phase transitions represented by orange lines indicate ordinary first order transitions, namely at these lines there is an energy crossing between two states but both of these states remain metastable in the immediate vicinity of these lines.

{\it \textcolor{blue}{Discussion.}}
We have studied the ground states of spontaneous symmetry broken integer quantum Hall states in the $N=1$ LL of graphene. We have constructed a general model consistent with the lattice symmetries of graphene that describes the short-range corrections to the Coulomb interaction. Based on this model we studied the ground states at integer fillings. We have found several important qualitative differences with respect to the $N=0$ LL. First, we showed that when a single component of the $N=1$ LL is filled $(\tilde{\nu}=1)$, our model can lift the degeneracy to select the ground states, in contrast to the $N=0$ LL where states remain undetermined. Moreover, interestingly, among the possible competing states at $\tilde{\nu}=1$, we find that spin-valley entangled phases can appear. On the other hand, when two components are filled ($ \tilde{\nu}=2 $), we have found a qualitatively new state that is absent in the $N=0$ LL, which features a combination of Kekul\'{e} and Antiferromagnet character and that we have termed the Kekul\'{e}-AF state.

We have shown that the related model for the $N=1$ LL that appeared in Ref.~\cite{yang2021experimental} is missing terms that are allowed by symmetry and is a special case of our model~\eqref{proj model}. In particular Ref.~\cite{yang2021experimental} is missing the inter-sublattice scattering interactions that appear in Eqs.~\eqref{Asymmetry}. By taking the parameters from Ref.~\cite{yang2021experimental}, we find that graphene will be near the phase boundary separating the CDW and AF states. However, this prediction should be taken carefully because of the aforementioned absence of A-B scattering processes in the model of Ref.~\cite{yang2021experimental}. These processes are known to be crucial in the $N=0$ LL, because they give rise to “$g_\perp$” interaction in Eq.~\eqref{strengths N=0}  that ultimately is needed to stabilize the AF or Kekul\'{e} states that are reported in experiments  ~\cite{li2019scanning,liu2022visualizing,coissard2022imaging,amet2015composite,yang2021experimental}. We see no reason why these inter-sublattice scattering terms would be negligible in the higher Landau levels. We hope our study stimulates future experiments to better narrow down the states and parameters realized in the $N=1$ LL of graphene.

\bibliography{ref}

\clearpage

\renewcommand{\theequation}{S-\arabic{equation}}
\renewcommand{\thefigure}{S-\arabic{figure}}
\renewcommand{\thetable}{S-\Roman{table}}
\makeatletter
\renewcommand\@biblabel[1]{S#1.}
\setcounter{equation}{0}
\setcounter{figure}{0}

\onecolumngrid

\begin{center}
\textbf{\large Supplemental Material: Theory of broken symmetry quantum Hall states in the $N=1$ Landau level of Graphene}
\end{center}

\section{S-I: Review of the model for graphene}
\par The tight binding Hamiltonian of graphene is :  
\begin{equation}
\begin{split}
H&=\sum_{\mathbf{R},\mathbf{R'},\sigma,\sigma'}t(\mathbf{R}+\sigma \mathbf{a_1} - \mathbf{R'}-\sigma' \mathbf{a_1}) \ket{\mathbf{R},\sigma}\bra{\mathbf{R'},\sigma'} \\
& = t \sum_{\mathbf{k}} \begin{pmatrix}
0 & M(\mathbf{k}) \\
M^{*}(\mathbf{k}) & 0 
\end{pmatrix}
\end{split}
\end{equation}
in the ordered basis $\ket{\mathbf{k} A}, \ \ket{\mathbf{k} B}$ with $\ket{\mathbf{k},\sigma}=\sum_{\mathbf{R}} e^{-i \mathbf{k} \mathbf{R}} \ket{\mathbf{R},\sigma}$ and $M(\mathbf{k})=e^{-i \mathbf{k} \mathbf{R}_1}+e^{-i \mathbf{k} \mathbf{R}_2}+e^{-i \mathbf{k} (\mathbf{R}_1+\mathbf{R}_2)}$. $\mathbf{R}$ labels the unit cell and $\sigma=\pm$ the sublattice. $\tau_i$ are the pauli matrices acting in valley space and $\sigma_i$ in sublattice space. Upon linearizing $M$, this gives the Dirac Hamiltonian : 
\begin{equation} \label{DiracS}
\begin{split}
\mathcal{H}_{D} &= \frac{\sqrt{3}}{2}Rt (\tau_z p_x \sigma_x+p_y \sigma_y) \\
\mathcal{H}_{D}&= v_F (\tau_z p_x \sigma_x+p_y \sigma_y)
\end{split}
\end{equation}
with $R=|\mathbf{R}_1|=|\mathbf{R}_2|$.
\begin{figure}[h!]
     \centering  
     \includegraphics[width=0.9\textwidth]{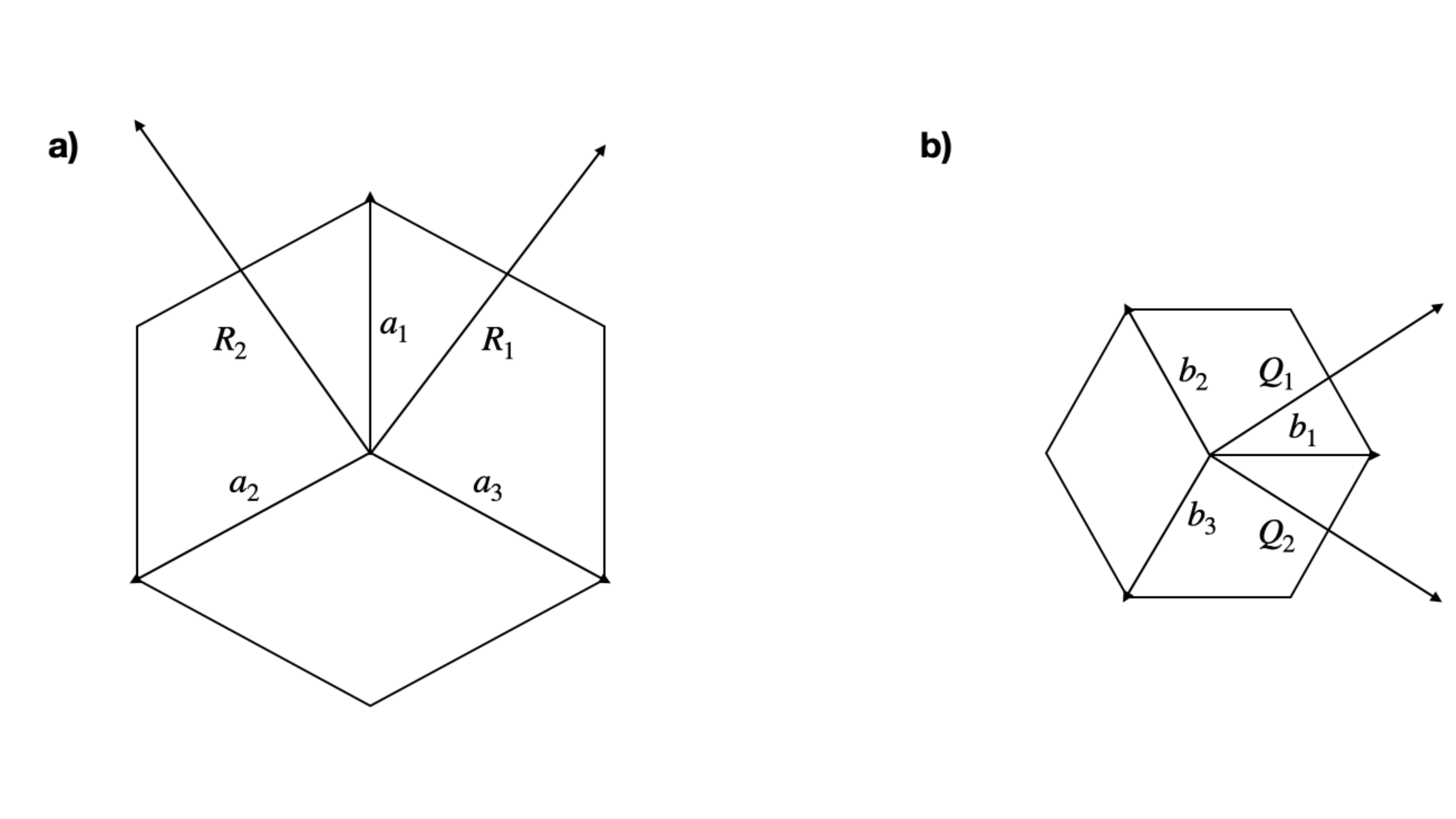}
     \caption{a) Unit cell of graphene. b) Brillouin zone.}
     \label{LatticeS}
\end{figure}
\par In a similar fashion, we can construct a model for graphene which describes the short range two-body interaction anisotropies appearing at the lattice scale of graphene. These can be written as : 
\begin{equation} \label{GenericS}
\mathcal{H}_A=\sum_{i<j}\{\sum_{\tau_1,...,\tau_4,\sigma_1,...,\sigma_4} V_{\tau_1,...,\tau_4 \linebreak \sigma_1,...,\sigma_4}\ket{\tau_{1}\sigma_1;\tau_{2}\sigma_2}_{ij}\bra{\tau_{3}\sigma_3;\tau_{4}\sigma_4}_{ij}\} \delta(r_i-r_j)
\end{equation} 
Here $\ket{\tau\sigma;\tau' \sigma'}$ label the two body hardcore eigenstates of Eq.~\eqref{Dirac} with $\tau=\pm 1 \  (\sigma= \pm 1)$ corresponding to $\tau= K,K' \  (\sigma=A,B)$ . Eq.~\eqref{GenericS} contains $2^8$ strengths which are subject to symmetry contraints. 
\subsection{General symmetries}
Because of the $SU_{s}(2)$ in spin space, the interactions can be decomposed into a singlet and triplet component : 
\begin{equation} \label{initialS}
\mathcal{H}_A=\mathcal{H}^{s}_A \bigoplus \mathcal{H}^{t}_{A}
\end{equation}
If we label the single particle states with a super-spin $\ket{N_{s}}=\ket{\tau_{1}\sigma_1}$ which takes $4$ values, we see that there are $6$ distinct spin triplet anti-symmetrized two body states consistent with this : 
\begin{equation}
\ket{\lambda}^{t}=\frac{1}{\sqrt{2}} (\ket{\tau_{1}\sigma_1;\tau_{2}\sigma_2}-\ket{\tau_{2}\sigma_2;\tau_{1}\sigma_1})
\end{equation} 
So we write Eq.~\eqref{GenericS} in the form : 
\begin{equation} \label{Generic1S}
\mathcal{H}^{t}_A=\sum_{i<j}\sum_{\lambda, \lambda'} V^{t}_{\lambda,\lambda'}\ket{\lambda}^{t}\bra{\lambda'}^{t} \delta(r_i-r_j)
\end{equation}

For the singlet, $10$ spin symmetric states exist : 
\begin{equation}
\ket{\lambda}^{s}=\frac{1}{\sqrt{2}} (\ket{\tau_{1}\sigma_1;\tau_{2}\sigma_2}+\ket{\tau_{2}\sigma_2;\tau_{1}\sigma_1})
\end{equation} 
and the interactions in this subspace can be written : 
\begin{equation} \label{Generic1S}
\mathcal{H}^{s}_A=\sum_{i<j}\sum_{\lambda, \lambda'} V^{s}_{\lambda,\lambda'}\ket{\lambda}^{s}\bra{\lambda'}^{s} \delta(r_i-r_j)
\end{equation}
Because of hermiticity, we always have : 
\begin{equation}
V_{\lambda, \lambda'}=V^{*}_{\lambda', \lambda}
\end{equation}
\subsection{Lattice symmetries of graphene}
Below we derive the action of the symmetry operations on the hardcore states $\ket{\tau, \sigma}$ introduced in Eqs.~\eqref{symm ops}. For the definition of the auxilary vectors used see Fig.~\ref{LatticeS}.
\begin{itemize}
    \item $C_6$. 
      \begin{equation}
          \begin{split}
            &C_{6}\ket{\mathbf{R},\sigma}= \ket{C_{6}\mathbf{R}+\sigma C_{6} \mathbf{a}_1}=\ket{C_6 \mathbf{R}+\sigma \mathbf{R}_{2},-\sigma} \\
            \rightarrow ~ & C_6 \ket{\mathbf{k},\sigma}=e^{-i \sigma (C_6\mathbf{k})\mathbf{R}_{2}} \\
             \rightarrow ~ &  C_6 \ket{\tau \mathbf{b}_1,\sigma}=Z^{-\tau \sigma} \ket{-\tau \mathbf{b}_1,-\sigma} \\
          \end{split}
      \end{equation}
      since $-\mathbf{b}_3=-\mathbf{b}_1+\mathbf{Q}_1\equiv -\mathbf{b}_1$, with $Z=e^{i 2 \pi/3}$.
    \item $M_{x}$.  $M_x \ket{\tau \mathbf{b}_1,\sigma}= \ket{\tau \mathbf{b}_1,-\sigma}$
    
    \item $M_{y}$. $M_y \ket{\tau \mathbf{b}_1,\sigma}= \ket{-\tau \mathbf{b}_1,\sigma}$
    \item $T_{\mathbf{R}_i}$. 
    \begin{equation}
    \begin{split}
    &T_{\mathbf{R}_i}\ket{\mathbf{R},\sigma}=\ket{\mathbf{R}+\mathbf{R}_i,\sigma} \rightarrow T_{\mathbf{R}_i}\ket{\mathbf{k},\sigma}=e^{i\mathbf{k} \mathbf{R}_i}\ket{\mathbf{k},\sigma}\\
    \rightarrow ~ & T_{\mathbf{R}_i}\ket{\tau,\sigma}=Z^{\pm \tau} \ket{\tau,\sigma} \\
    \end{split}
    \end{equation}
\end{itemize}

\subsection{Symmetry reduced model}
By using the aforementioned symmetries, the model in Eq.~\eqref{GenericS} can be block-diagonalized and reduced to a total of $9$ independent real parameters. This yields :
\begin{equation} \label{paulisS}
\begin{split}
\mathcal{H}^{A}&= \sum_{i<j}\Bigg\{F_{00}(\tau^{i}_{0} \tau^{j}_{0})(\sigma^{i}_{0} \sigma^{j}_{0})+F_{z \perp}\bigg((\tau^{i}_{0} \tau^{j}_{0})(\sigma^{i}_{x} \sigma^{j}_{x}) +(\tau^{i}_{z} \tau^{j}_{z})(\sigma^{i}_{y} \sigma^{j}_{y})\bigg) \\
& +F_{0 \perp}\bigg((\tau^{i}_{0} \tau^{j}_{0})(\sigma^{i}_{y} \sigma^{j}_{y}) +(\tau^{i}_{z} \tau^{j}_{z})(\sigma^{i}_{x} \sigma^{j}_{x})\bigg) +F_{zz}(\tau^{i}_{0} \tau^{j}_{0})(\sigma^{i}_{z} \sigma^{j}_{z})\\
& +\frac{F_{\perp \perp}}{2}\tau^{i}_{\perp} \tau^{\perp}_{\perp}\bigg(\sigma^{i}_{0} \sigma^{j}_{0} +\sigma^{i}_{z} \sigma^{j}_{z}\bigg)\\
&+\frac{F_{\perp z}}{2}\bigg((\tau^{i}_{+} \tau^{j}_{-}+\tau^{i}_{-} \tau^{j}_{+})(\sigma^{i}_{x} \sigma^{j}_{x}) \bigg) +\frac{F_{\perp 0}}{2} \bigg((\tau^{i}_{+} \tau^{j}_{-}+\tau^{i}_{-} \tau^{j}_{+})(\sigma^{i}_{y} \sigma^{j}_{y}) \bigg)\\
&+ F_{z0}(\tau^{i}_{z} \tau^{j}_{z})(\sigma^{i}_{0} \sigma^{j}_{0})+F_{0z}(\tau^{i}_{z} \tau^{j}_{z})(\sigma^{i}_{z} \sigma^{j}_{z})\Bigg\}\delta(r_i-r_j)\\
\end{split}
\end{equation}
We note that the model introduced in Ref.~\cite{aleiner2007spontaneous} is written in the basis of :
\begin{equation}
\begin{pmatrix}
 \ket{KA} \\
 \ket{KB} \\
 \ket{K'B}\\
 -\ket{K'A}\\
\end{pmatrix}
\end{equation}
whereas ours in Eq.~\eqref{paulisS} in the basis of:
\begin{equation}
\begin{pmatrix}
 \ket{KA} \\
 \ket{KB} \\
 \ket{K'A}\\
 \ket{K'B}\\
\end{pmatrix}
\end{equation}
\section{S-II: Projected model into the $N$th LL} \label{Projected model}
The Dirac Hamiltonian in a magnetic field can be derived from the substitution $ \mathbf{p} \rightarrow \boldsymbol{\Pi}=\mathbf{p}+e\mathbf{A}$, with $[\Pi_x, \Pi_y]=-i$. We have : 
\begin{equation}
\begin{split}
H&= v_F (\tau_z \Pi_x \sigma_x+\Pi_y \sigma_y) 
\end{split}
\end{equation}
We flip the basis in valley $K'$ : 
\begin{equation}
\psi_{\tau=K}= \begin{pmatrix}
\ket{\tau=K, A}  \\
\ket{\tau=K, B}
\end{pmatrix}
\end{equation}
\begin{equation}
\psi_{\tau=K'}= \begin{pmatrix}
\ket{\tau=K', B}  \\
\ket{\tau=K', A}
\end{pmatrix}
\end{equation}
and the hamiltonian becomes : 
\begin{equation}
H= \tau_z v_F (\Pi_x \sigma_x+\Pi_y \sigma_y) 
\end{equation}
By defining $\hat{a}=\frac{1}{\sqrt{2}}(\Pi_x-i\Pi_y)$ and $\hat{a}^{\dagger}=\frac{1}{\sqrt{2}}(\Pi_x+i\Pi_y)$ with $[\hat{a},\hat{a}^{\dagger}]=1$, the hamiltonian becomes:
\begin{equation}
H= \tau_z \sqrt{2} v_F \begin{pmatrix}
0 &  \hat{a}\\
\hat{a}^{\dagger} & 0 \\
\end{pmatrix}
\end{equation}
The eigenstates of the Dirac Hamiltonian in the $N$th LL have a definite valley number and can be written as : 
\begin{equation} \label{eigenstatesS}
    \begin{split}
        \ket{K} & =\frac{1}{\sqrt{2}} \{ \ket{n=N-1, A}+\ket{n=N,B} \} \\
        \ket{K'}&=\frac{1}{\sqrt{2}} \{ \ket{n=N-1, B}-\ket{n=N,A} \} \\
    \end{split}
\end{equation}
The model for the short range corrections to the Coulombs in the $N=1$ LL, $\mathcal{H}^{N}_A$  can be found by projecting $\mathcal{H}_A$ in Eq.~\eqref{paulisS} to the states in Eq.~\eqref{eigenstatesS} , $\ket{\tau}$ , with $\tau=K,K' \  (\bar{\tau}=K',K)$: 
\begin{equation} \label{projectionS}
   \begin{split}
   \mathcal{H}^{N}_A & =\sum_{i<j} \sum_{\tau_1, \tau_2, \tau_3, \tau_4} \braket{\tau_1, \tau_2|_{ij}\mathcal{H}^{ij}_A|\tau_3, \tau_4}_{ij} \ket{\tau_1, \tau_2}_{ij} \bra{\tau_3, \tau_4}_{ij} \\
   & \equiv \sum_{i<j} \sum_{\tau_1, \tau_2, \tau_3, \tau_4} \mathcal{H}^{ij}_{A, \tau_1 ... \tau_4} \ket{\tau_1, \tau_2}_{ij} \bra{\tau_3, \tau_4}_{ij}
   \end{split}
\end{equation}
, with $\mathcal{H}^{ij}_A$ the interaction between particles $i,j$ obtained from Eq.~\eqref{paulisS}. Due to the $U_{v}(1)$ of $\mathcal{H}_A$, Eq.~\eqref{projectionS} reduces to :
\begin{equation} \label{general form of the modelS}
\begin{split}
    \mathcal{H}^{N}_A & = \sum_{i<j}\sum_{\tau_1, \tau_2} \{ \mathcal{H}^{ij}_{A, \tau_1 \tau_2 \tau_1 \tau_2} \ket{\tau_1, \tau_2}_{ij} \bra{\tau_1, \tau_2}_{ij} + \mathcal{H}^{ij}_{A, \tau_1 \tau_2 \bar{\tau_1} \bar{\tau_2}} \ket{\tau_1, \tau_2}_{ij} \bra{\bar{\tau_1}, \bar{\tau_2}}_{ij} \delta_{\tau_1, - \tau_2} \} \\
    &=\sum_{i<j} \{ V^{N}_{0}(r_{ij}) \tau^{i}_{0} \tau^{j}_{0} + V^{N}_{z}(r_{ij}) \tau^{i}_{z}\tau^{j}_{z} +V^{N}_{\perp}(r_{ij})\tau^{i}_{\perp}\tau^{j}_{\perp} \} \\
    \end{split}
\end{equation}
Due to $C_6$ symmetry, $\mathcal{H}^{KKKK}_{A}=\mathcal{H}^{K'K'K'K'}_{A} , \ \mathcal{H}^{K K' K K'}_{A}= \mathcal{H}^{K' K K' K}_{A}$, we have: 
\begin{equation}
  \begin{split}
      V_{0}&=\frac{1}{2} ( \mathcal{H}^{KKKK}_{A}+ \mathcal{H}^{KK'KK'}_{A})\\
      V_{z}&=\frac{1}{2} ( \mathcal{H}^{KKKK}_{A}- \mathcal{H}^{KK'KK'}_{A})\\
      V_{\perp}&= \mathcal{H}^{KK'KK'}_{A}=\mathcal{H}^{K'KK'K}_{A}
  \end{split}
\end{equation}
The $V_{0}$ term is negligible compared to the $SU(4)$ symmetric Coulombs and will be omitted. Therefore, we reach to the general form for the $N$th LL in the main text. 

\par More concretely, by writing the position operators in terms of the guiding center operators $R_i=r_i+\epsilon_{ij} \Pi_j$, the matrix elements $\mathcal{H}^{ij}_{A, \tau_1 ... \tau_4}$ can be written as : 
\begin{equation}
  \mathcal{H}^{ij}_{A, \tau_1 ... \tau_4} =\sum_{\mathbf{q}}\sum_{a,\beta}e^{-i \mathbf{q} (\mathbf{R}_i-\mathbf{R}_j)}F_{a \beta}  \braket{\tau_1|\hat{F}_{i}(\mathbf{q}) \tau_{a}\sigma_{\beta}| \tau_3} \braket{\tau_2|\hat{F}_{j}(\mathbf{-q})\tau_{a} \sigma_{\beta}| \tau_4} 
\end{equation}
with $\hat{F}_{i}(\mathbf{q})=e^{i \mathbf{q} \cdot (\hat{z}\times \boldsymbol{\Pi})}$. By evaluting these matrix elements, we reach to : 
\begin{equation}
\begin{split}
V^{N}_{z}(\mathbf{q}) & = \frac{1}{4} \{ (|F_{N-1,N-1}(\mathbf{q})|^2+|F_{N,N}(\mathbf{q})|^2) (F_{z0}+F_{zz}) \\
&+ (F_{N-1,N-1}(\mathbf{q}) F_{N.N}(-\mathbf{q})+F_{N,N}(\mathbf{q}) F_{N-1,N-1}(-\mathbf{q})) (F_{z0}-F_{zz})+|F_{N-1,N}(\mathbf{q})|^2F_{z \perp}\} \\
V^{N}_{\perp}(\mathbf{q}) & =  \frac{1}{4} \{  F_{\perp \perp}|F_{N-1,N}(\mathbf{q})|^2+\frac{1}{4}(F_{\perp 0}-F_{\perp z})F_{N-1,N-1}(\mathbf{q})F_{N,N}(-\mathbf{q})  \\
&+\frac{1}{8} (F_{\perp z}+F_{\perp 0}) (|F_{N-1,N-1}(\mathbf{q})|^2+F_{N,N}(\mathbf{q})|^2) \} \\
\end{split}
\end{equation}
with $F_{n,n'}=\braket{n| e^{i \mathbf{q} \cdot (\hat{z}\times \mathbf{\Pi})}|n'}$ being the form factors in the LL $n,n'$.
\par In the $N=1$ LL, the strengths $V_{z,\perp}$, take the form : 
\begin{equation} \label{strengthsN=1S}
V_{z,\perp}(r_{ij}) = \sum_{n=0}^{2}g^{z,\perp}_n \nabla^{2n} \delta(r_{ij})
\end{equation}
where we have used the form factors :
\begin{equation}
\begin{split}
&F_{0,0}(\mathbf{q})=e^{-\frac{|q|^2}{4}} \\
&F_{1,0}(\mathbf{q})=-\frac{iq^{*}}{\sqrt{2}}e^{-\frac{|q|^2}{4}} \\
&F_{0,1}(\mathbf{q})=-\frac{iq}{\sqrt{2}}e^{-\frac{|q|^2}{4}} \\
&F_{1,1}(\mathbf{q})=(1-\frac{|q|^2}{2})e^{-\frac{|q|^2}{4}} \\
\end{split}
\end{equation}
with $q=q_x-iq_y$. The relation of the parameters of the projected model to the original one is : 
\begin{equation} \label{eqs to microscopic paramsS}
\begin{split}
    &g^{z}_{0}=F_{z0} \\
    &g^{z}_{1}=\frac{F_{z\perp}}{8}+ \frac{F_{z0}}{2}\\
    &g^{z}_{2}=\frac{F_{zz}}{16}+\frac{F_{z0}}{16}\\
    &g^{\perp}_{0}=\frac{F_{\perp 0}}{8} \\
    &g^{\perp}_{1}=\frac{F_{\perp 0}}{16}+ \frac{F_{\perp \perp}}{4}\\
    &g^{\perp}_{2}=\frac{1}{128}(F_{\perp z}+F_{\perp 0})\\
\end{split}
\end{equation}
For the $N=0$ LL the parameters of the projected model are : 
\begin{equation} \label{strengthsN=0S}
\begin{split}
& g_{\perp}=F_{\perp z}+F_{\perp 0} \\
& g_{z}=F_{zz}+F_{z0} \\
\end{split}
\end{equation}

\section{S-III: Hartree-Fock theory} \label{Parameters}
In first quantization the projected hamiltonian both in the $N=0$ and $N=1$ LL have the following form : 
\begin{equation}
V^{P}=\sum_{a} \sum_{i<j} \sum_{\mathbf{q}} e^{i \mathbf{q} \cdot (\mathbf{R}_i-\mathbf{R}_j)} V_{a} (\mathbf{q}) \tau^i_{a}\tau^j_{a}
\end{equation}
In second quantization this becomes : 
\begin{equation}
V^{P}=\frac{1}{2A}\sum_{a} \sum_{X_1,...X_4,s_1,s_2,\tau_1,...\tau_4} \sum_{\mathbf{q}} V_{a} (\mathbf{q}) \rho_{X_1;X_4}(\mathbf{q}) \rho_{X_2;X_3}(-\mathbf{q}) \tau^{\tau_1 \tau_4}_{a}\tau^{\tau_2 \tau_3}_{a} c^{\dagger}_{X_1 \tau_1 s_1}c^{\dagger}_{X_2 \tau_2 s_2} c_{X_3 \tau_3 s_2} c_{X_4 \tau_4 s_1}
\end{equation}
with $ \rho_{X_1;X_2}(\mathbf{q})=\delta_{X_1,X_2+q_y} e^{i \frac{q_x (X_1+X_2)}{2}}$.  
\par We search for the mean field energy functional for translational invariant states, parametrized by : $\braket{c^{\dagger}_{X_1 \tau_1 s_1} c_{X_2 \tau_2 s_2}}=P^{s_1 s_2}_{\tau_1 \tau_2} \delta_{X_1,X_2}$. We find : 
\begin{equation}
E_{HF}[P]=\frac{A}{8 \pi^2}\sum_{a} \Big \{ V_{a}(\mathbf{0}) Tr \{ T_{a} P\} Tr \{ T_{a} P\} -\frac{1}{2 \pi}( \iint dq_x dq_y V_{a}(q_x,q_y)) Tr \{ T_{a} P  T_{a} P\} \Big\}
\end{equation}
\par For the $N=0$ LL we get : 
\begin{equation}
\begin{split}
E_{HF}[P] &=\frac{N^2_{\phi}}{2 A}\sum_{a}g_{a} \Big \{ Tr \{ T_{a} P\} Tr \{ T_{a} P\} -Tr \{ T_{a} P  T_{a} P\}  \Big \} \\
&=\frac{1}{8 \pi^2}\sum_{a} g_{a} \int d\mathbf{r} \Big \{ Tr \{ T_{a} P\} Tr \{ T_{a} P\} -Tr \{ T_{a} P  T_{a} P\}  \Big \} \\
\end{split}
\end{equation}
where $g_{a}$ the interaction strengths in the $N=0$ LL in Eq.~\eqref{strengthsN=0S} for $a=\perp, z $.
\par For the $N=1$ LL we get : 
\begin{equation} \label{HF energyS}
\begin{split}
E_{HF}[P] &=\frac{N^2_{\phi}}{2 A}\Big \{ u^{H}_{z} Tr \{ T_{z} P\} Tr \{ T_{z} P\} -u^{X}_{z}Tr \{ T_{z} P  T_{z} P\} \\
&+u^{H}_{\perp} \big (Tr \{ T_{x} P\} Tr \{ T_{x} P\} + Tr \{ T_{y} P\} Tr \{ T_{y} P\} \big ) \\
&-u^{X}_{\perp}(Tr \{ T_{x} P  T_{x} P\}+Tr \{ T_{y} P  T_{y} P\}) \Big \} \\
\end{split}
\end{equation}
with 
\begin{equation}
\begin{split}
& u^{H}_{z}= F_{z0} \\
& u^{X}_{z}= -\frac{F_{z \perp}}{4} +F_{zz}+F_{z0} \\
 & u^{H}_{\perp}=\frac{F_{\perp 0}}{8} \\
& u^{X}_{\perp}=-\frac{F_{\perp \perp}}{2}+\frac{F_{\perp z}}{16}+\frac{F_{\perp 0}}{16} \\ 
\end{split}
\end{equation}
\section{S-IV: Comparison with the model of Ref.~\cite{yang2021experimental}}
The model proposed in Eq.(S21) of Ref.~\cite{yang2021experimental} can be recasted into the following more convenient form before projection, 
\begin{equation} \label{Zaletel's modelS}
    \mathcal{V}= \sum_{i<j} \{V_{1}\tau^{i}_{0} \tau^{j}_{0} (\sigma^{i}_{0}\sigma^{j}_{0}+\sigma^{i}_{z}\sigma^{j}_{z})+V_{2} \tau^{i}_{\perp} \tau^{j}_{\perp} (\sigma^{i}_{0}\sigma^{j}_{0}+\sigma^{i}_{z}\sigma^{j}_{z})\} \delta(\mathbf{r}_i-\mathbf{r}_j)
\end{equation}
Notice that while the above model contains inter-valley scattering terms, it does not contain inter-sublattice scattering terms. This model is a special case of the Aleiner, Kharzeev and Tsvelik, in which the only non-vanishing parameters are $F_{zz}, \ F_{\perp \perp} \neq 0$ in Eq.~\eqref{paulisS}.
Upon projection of Eq.~\eqref{Zaletel's modelS} we find that,
\begin{equation}
    \begin{split}
        & g^{z}_{0,1}=g^{\perp}_{0,2}=0 \\
        & g^{z}_{2}= \frac{F_{zz}}{16} , \ g^{\perp}_{1}=\frac{F_{\perp \perp}}{4} \\
    \end{split}
\end{equation}
, leading to Eq.(1) of Ref.~\cite{yang2021experimental}. Moreover, Ref.~\cite{yang2021experimental} also estimated that the above constants $g^{z}_{2},g^{\perp}_{1}$ are positive. Therefore this leads to the following values and signs of the parameters of the HF functional:
\begin{equation}
    \begin{split}
      & u^{H}_{z, \perp}=0 \\
      & \Delta_z<0, \ \Delta_{\perp}>0, \ \Delta_z<\Delta_{\perp} \\
    \end{split}
\end{equation}
\section{S-V-A): Ground states at quarter-filling ($\tilde{\nu}=1$)}\label{S-III-a)}
Let's label the occupied state by $\ket{\chi_{i}}$. Then we have $P=\ket{\chi_{i}}\bra{\chi_{i}}$. So, 
\begin{equation}
(Tr \{ P T_a \})^2=Tr \{ P T_a P T_a\}=(\braket{\chi_i|T_a|\chi_i})^2
\end{equation}
since only one state contributes to the trace. 
Then the HF energy would be : 
\begin{equation} \label{quarterS}
\begin{split}
& E^{N=0}_{HF}=0 \\
& E^{N=1}_{HF}=\frac{N^2_{\phi}}{2 A} \Big \{ (u^{H}_z-u^{X}_{z}) \braket{\chi_1|T_z|\chi_1}^2+ \\
&  (u^{H}_{\perp}-u^{X}_{\perp}) (\braket{\chi_1|T_x|\chi_1}^2+\braket{\chi_1|T_y|\chi_1}^2) \Big \}\\
&=\frac{N^2_{\phi}}{2 A} \Big \{ \Delta_z \braket{\chi_1|T_z|\chi_1}^2+ \\
&  \Delta_{\perp} (\braket{\chi_1|T_x|\chi_1}^2+\braket{\chi_1|T_y|\chi_1}^2) \Big \}\\
\end{split}
\end{equation}

\section{S-V-B): Ground states at half-filling ($\tilde{\nu}=2$)}\label{S-III-b)}
We first consider the disentangled, valley and spin active,  states in Eqs.~\eqref{valley active states}, ~\eqref{spin active states}. The HF functional for the valley active states is, 
\begin{equation} \label{disentangled valleyS}
\begin{split}
E_{HF}[P_{\boldsymbol{\eta}}] &=\frac{N^2_{\phi}}{2 A}\Big \{ u^{H}_{z} (\eta^z_{1}+\eta^z_{2})^2 -u^{X}_{z} ((\eta^z_{1})^2+(\eta^z_{2})^2) \\
&+u^{H}_{\perp} \big ( (\eta^x_{1}+\eta^x_{2})^2 + (\eta^y_{1}+\eta^y_{2})^2 \big ) \\
&-u^{X}_{\perp}((\eta^x_{1})^2+(\eta^x_{2})^2 +(\eta^y_{1})^2+(\eta^y_{2})^2\}) \Big \} \\
\end{split}
\end{equation}
and for the spin active, 
\begin{equation} \label{disentangled spinS}
\begin{split}
E_{HF}[P_{\mathbf{s}}]&= \frac{N^2_{\phi}}{2 A} \Big \{ -(2 u^{X}_{\perp}+u^{X}_{z})(1+ \mathbf{s}_1 \cdot \mathbf{s}_2) - \\
&(u^{X}_{\perp} (n_{\perp})^2+u^{X}_{z}(n_z)^2)(1- \mathbf{s}_1 \cdot \mathbf{s}_2) \Big \} \\
\end{split}
\end{equation}
It is easy to recover the $N=0$ LL functionals by setting $u_{\perp, z}^{H}=u_{\perp, z}^{X}$. We note that for the spin active states the Hartree energy vanishes, rendering the HF functional for the spin active states the same as in the $N=0$ LL. In Table ~\ref{TableofEnergiesS}, the energies of the different states can be found.
\begin{table}[h!]
\begin{tabular}{ |p{4cm}||p{4cm}||p{4cm}|} 
 \hline
 \multicolumn{3}{|c|}{States appearing in the $\tilde{\nu}=2$} \\
 \hline
States & Wavefunctions & Energies \\
 \hline
 CDW (Charge density wave)  & $\ket{F}_1=\ket{\eta_z} \ket{\mathbf{s}}, \ \ket{F}_2=\ket{\eta_z} \ket{-\mathbf{s}}$ & $\mathcal{E}_{HF}= 2 \Delta_z+2 u^{H}_{z}$\\
 KD (Kekul\'{e} distortion) & $ \ket{F}_1 = \ket{\eta_{\perp}} \ket{\mathbf{s}}, \ \ket{F}_2=\ket{\eta_{\perp}} \ket{-\mathbf{s}}$ & $\mathcal{E}_{HF}=2 \Delta_{\perp}+2 u^{H}_{\perp} $ \\
 FM (Ferromagnet)& $\ket{F}_1=\ket{\eta_z} \ket{\mathbf{s}}, \ \ket{F}_2=\ket{-\eta_z} \ket{\mathbf{s}}$ & $ \mathcal{E}_{HF}= 4 \Delta_{\perp}+2 \Delta_z - 4 u^{H}_{\perp}-2 u^{H}_{z}$\\
 AF  (Antiferromagnet)  & $\ket{F}_1=\ket{\eta_z} \ket{\mathbf{s}}, \ \ket{F}_2=\ket{-\eta_z} \ket{-\mathbf{s}}$ & $\mathcal{E}_{HF}=  2 \Delta_z-2u^{H}_{z}$\\
 KD-AF (Kekul\'{e} antiferromagnet)& $\ket{F}_1=\ket{\eta_{\perp}} \ket{\mathbf{s}}, \ \ket{F}_2=\ket{-\eta_{\perp}} \ket{-\mathbf{s}}$ & $\mathcal{E}_{HF}= 2 \Delta_{\perp}-2 u^{H}_{\perp}$\\
 \hline
\end{tabular}
\caption{Table representing the states appearing in the phase diagram at $\tilde{\nu}=2$, their wavefunctions and their HF energies.}
\label{TableofEnergiesS}
\end{table}

\begin{figure*}[h!] 
\centering
\includegraphics[width=0.8\textwidth]{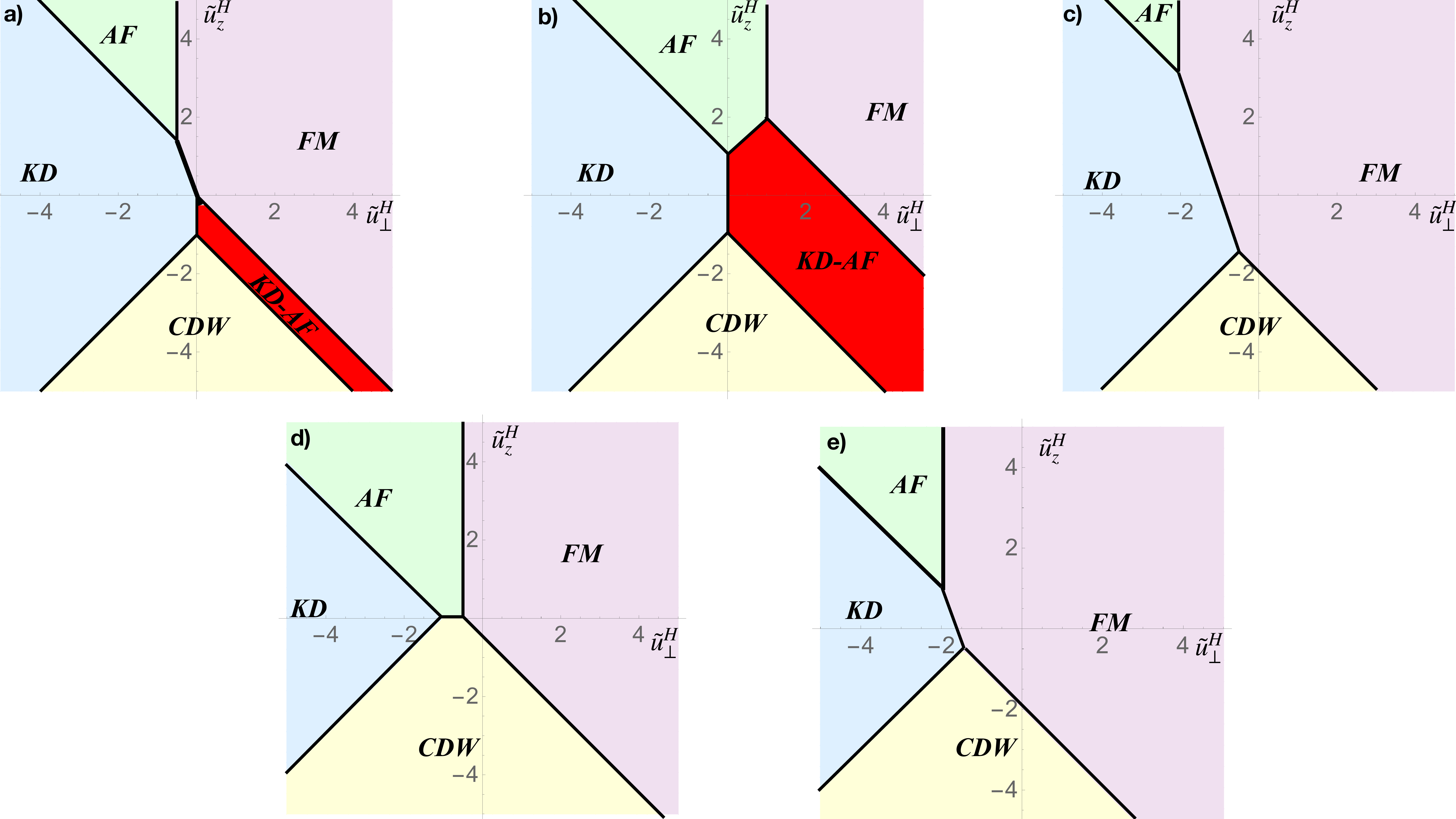}   
\caption{a)Phase diagram for $\Delta_z>0 , \ \Delta_{\perp}<0, \ \Delta_z >\Delta_{\perp}$ for $\Delta_z/\Delta_{\perp}=-1$. b)Phase diagram for $\Delta_{z,\perp}>0 , \ \Delta_z >\Delta_{\perp}$ for $\Delta_z/\Delta_{\perp}=2$. c)Phase diagram for $\Delta_{z,\perp}<0 , \ \Delta_z >\Delta_{\perp}$ for $\Delta_z/\Delta_{\perp}=1/2$ . d)Phase diagram for $\Delta_{z,\perp}<0, \ \Delta_{z}<\Delta_{\perp} , \ \Delta_z <2 \Delta_{\perp} $ for $\Delta_z/\Delta_{\perp}=3$. e)Phase diagram for $ \Delta_{z,\perp}<0, \ \Delta_{z}<\Delta_{\perp},   \ \Delta_z >2\Delta_{\perp}$ for $\Delta_z/\Delta_{\perp}=2/3$.}
\label{FIGS_SUPPLEMENTARYS}  
\end{figure*}

In addition to the phase diagrams in the main text, we obtain by comparing the HF energies of the states for the other possible cases of the values of $\Delta_{z, \perp}$ the phase diagrams in Figs.~\ref{FIGS_SUPPLEMENTARYS}.
\section{S-VI: Linear stability analysis in valley-spin disentangled sub-spaces} \label{Disentangled}
By expanding around the states with minimum energy up to quadratic terms, we are able to find the stability lines for spin-valley disentangled fluctuations. For the valley active states, by writing : 
\begin{equation}
    n^{z,\perp}_{1,2} \approx \pm \big( 1-\frac{1}{2}(n^{\perp,z}_{1,2})^2 \big)
\end{equation}
the HF energies up to quadratic fluctuations can be written as : 
\begin{equation}
    \mathcal{E}^i[\delta n_{z,\perp}]= \mathcal{C}^{i}+\frac{1}{2}\delta n^{T}_{z, \perp} \mathbb {K}^{i} \delta n_{z, \perp}
\end{equation}
with $i$ representing the state, i.e CDW, KD, KD-AF, AF , $\mathbb {K}^{i}$ the stability matrix, $\mathcal{C}^{i}$ constants and $\delta n_{z, \perp}=\begin{pmatrix}
\delta n^{1}_{z, \perp}\\
 \delta n^{2}_{z, \perp}
\end{pmatrix} \\$ the fluctuations around the ground state for the two components. 
\par
\begin{figure*}[h!]
\centering
\includegraphics[width=0.8\textwidth]{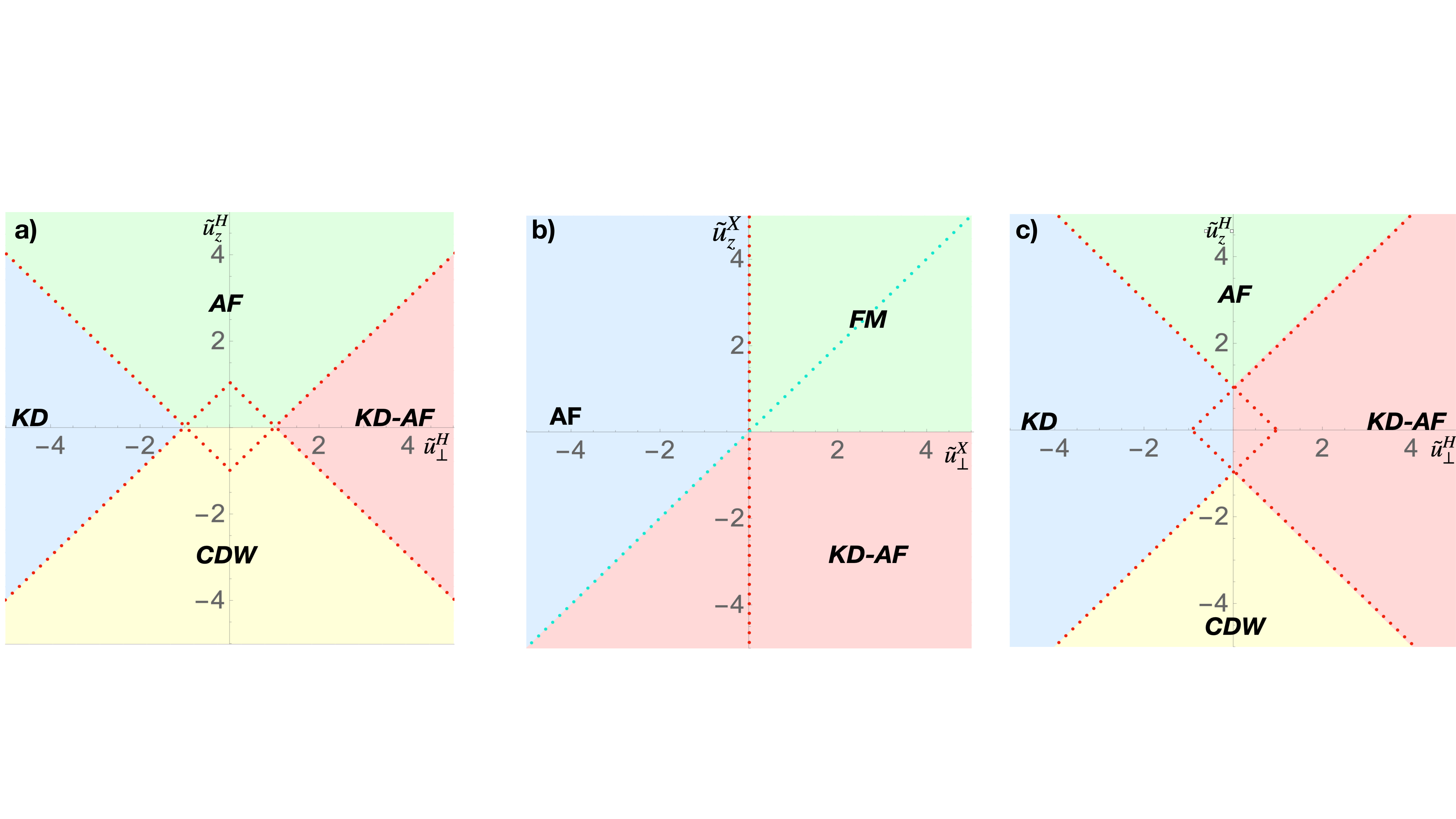}   \caption{a) Phase diagram (coloured regions in the phase diagram) and stability lines (dotted red lines) for the valley active states for $\Delta_z > \Delta_{\perp} $. b) Phase diagram (coloured regions in the phase diagram) and stability lines (dotted red lines) for the spin active states. The red dotted line represents the stability line of the FM, while the blue one of the KD-AF and KD. c) Phase diagram (coloured regions in the phase diagram) and stability lines (dotted red lines) for the valley active states for $\Delta_z < \Delta_{\perp} $.}
\label{Stability_analysis_figuresS}  
\end{figure*}

 The stability matrices are : 
\begin{equation}
\begin{split}
\mathbb{K}^{AF}&=
\begin{pmatrix}
2u^{H}_{z}\pm2& 2u^{H}_{\perp}\\
 2u^{H}_{\perp} &2u^{H}_{z}\pm 2
\end{pmatrix} \\
\mathbb{K}^{CDW}&=
\begin{pmatrix}
-2u^{H}_{z}\pm2& 2u^{H}_{\perp}\\
 2u^{H}_{\perp} &-2u^{H}_{z}\pm2
\end{pmatrix} \\
\mathbb{K}^{KD-AF}&=
\begin{pmatrix}
2u^{H}_{\perp}\mp 2& 2u^{H}_{z}\\
 2u^{H}_{z} & 2u^{H}_{\perp}\mp 2
\end{pmatrix} \\
\mathbb{K}^{KD}&=
\begin{pmatrix}
-2u^{H}_{\perp}\mp 2& 2u^{H}_{z}\\
 2u^{H}_{z} & -2u^{H}_{\perp}\mp 2
\end{pmatrix} \\
\end{split}
\end{equation}
, where the upper (the lower) sign corresponds to $\Delta_z< (>) \Delta_{\perp}$. For the spin active states, an analogous analysis yields, 
\begin{equation}
\begin{split}
    \mathcal{E}^{FM}[\delta \mathbf{s}_{1,2}]&=\mathcal{C}^{FM}-2u^{X}_{\perp} \delta \mathbf{s}_1 \cdot \delta \mathbf{s}_2 \\
    \mathcal{E}^{KD, KD-AF}[\delta \mathbf{s}_{1,2}]&=\mathcal{C}^{KD, KD-AF}-2u^{X}_{\perp} \delta \mathbf{s}_1 \cdot \delta \mathbf{s}_2 -2(u^{X}_{\perp}-u^{X}_{z})n^2_{\perp} \\
\end{split}
\end{equation}
Our results are presented in Figs.~\ref{Stability_analysis_figuresS}.

\section{S-VII: Linear stability analysis in valley-spin entangled spaces for the CDW and AF} \label{stability CDW AF}
The general HF functional for the spin-valley entangled states, 
\begin{equation} 
\begin{split}
\ket{F}_1&=\cos\frac{a_1}{2} \ket{\boldsymbol{\eta}}\ket{\mathbf{s}}+e^{i \beta_1} \sin \frac{a_1}{2} \ket{-\boldsymbol{\eta}} \ket{-\mathbf{s}} \\
\ket{F}_2&=\cos\frac{a_2}{2} \ket{\boldsymbol{\eta}}\ket{\mathbf{-s}}+e^{i \beta_2} \sin \frac{a_2}{2} \ket{-\boldsymbol{\eta}} \ket{\mathbf{s}}\\
\end{split}
\end{equation}

is : 
\begin{equation} \label{General HFS}
\begin{split}
&\mathcal{E}_{HF}= 2u^{H}_{z} \{ M_{z}^{P_1} M_{z}^{P_2}- |M_{z}^{P_{12}}|^2\} \\
&+2u^{H}_{\perp} \{ M_{x}^{P_1} M_{x}^{P_2}+ M_{y}^{P_1} M_{y}^{P_2}- |M_{x}^{P_{12}}|^2- |M_{y}^{P_{12}}|^2\} \\
&+\Delta_z \{ |M_{z}^{P_1}|^2 + |M_{z}^{P_2}|^2+ 2 |M_{z}^{P_{12}}|^2\} \\
&+\Delta_{\perp} \{ |M_{x}^{P_1}|^2 + |M_{y}^{P_2}|^2+ 2 |M_{x}^{P_{12}}|^2 \\
& +|M_{y}^{P_1}|^2 + |M_{y}^{P_2}|^2+ 2 |M_{y}^{P_{12}}|^2\} \\
\end{split}
\end{equation}
with $M_{a}^{P_{i(j)}}=\braket{F_i|\tau_a|F_j}$ and we can always write the unit vector $\boldsymbol{\eta}$ as ,
$\boldsymbol{\eta}= \begin{pmatrix}
  \sin \theta_p \cos \phi_p \\ 
  \sin \theta_p \sin \phi_p  \\
  \cos \theta_p 
\end{pmatrix}$.
\subsection{CDW}

By expanding around $a_1=a_2=\theta_p=0$ ~\eqref{General HFS} keeping up to quadratic terms, the energy functional for the CDW is around the minimum is : 
\begin{equation}
\begin{split}
\mathcal{E}_{CDW}&=(a_1^2+a_2^2)(\Delta_{\perp}-\Delta_z-u^{H}_{\perp}-u^{H}_{z}) \\
&+2 (\Delta_{\perp}-\Delta_z+u^{H}_{\perp}-u^{H}_{z}) \theta_p^2 \\ 
\end{split}
\end{equation}
So we find that the instability lines are:
\begin{equation}
\begin{split}
u^{H}_{z}&=-u^{H}_{\perp}+(\Delta_{\perp}-\Delta_z) \\
u^{H}_{z}&=u^{H}_{\perp}+(\Delta_{\perp}-\Delta_z) \\
\end{split}
\end{equation} 
\subsection{AF}
By expanding around $a_1=a_2=\theta_p=\frac{\pi}{2}$ and $\beta=\beta_1+\beta_2=0$, the energy is : 
\begin{equation}
\begin{split}
\mathcal{E}_{AF}&=(a_1^2+a_2^2)(3\Delta_{\perp}-\Delta_z-u^{H}_{\perp}+u^{H}_{z}) \\
&-(\Delta_{\perp}+\Delta_z-3u^{H}_{\perp}-u^{H}_{z})a_1 a_2\\
&+ (\Delta_{\perp}-\Delta_z-u^{H}_{\perp}+u^{H}_{z}) \beta^2 \\
&+4 (\Delta_{\perp}-\Delta_z-u^{H}_{\perp}+u^{H}_{z}) \theta_p^2 \\
\end{split}
\end{equation}
So we find that the instability lines are:
\begin{equation}
\begin{split}
u^{H}_{z}&=u^{H}_{\perp}+(\Delta_{z}-\Delta_{\perp}) \\
u^{H}_{\perp}&=\Delta_{\perp}\\
u^{H}_{z}&=-u^{H}_{\perp}+(\Delta_{z}-\Delta_{\perp}) \\
\end{split}
\end{equation} 
These are the same as the ones which occur from the non spin-valley entangled analysis.

\end{document}